\documentstyle[epsf,12pt]{article}
\newcommand{\bq}{\begin{equation}}
\newcommand{\eq}{\end{equation}}
\newcommand{\ba}{\begin{eqnarray}}
\newcommand{\ea}{\end{eqnarray}}
\newcommand{\nobody}{\rule{0ex}{1ex}}

\newcommand{\ra}{\rightarrow}
\begin{document}
\thispagestyle{empty}
\begin{flushright}
MPI-PhT/94-84\\ LMU-23/94\\ November 1994\vspace*{3cm}
\end{flushright}
\begin{center}
{\LARGE\bf Production Mechanisms for $B_c$ Mesons in Photon--Photon
Collisions\footnote
{Supported by the German Federal Ministry for Research
and Technology under contract No.~05~6MU93P.}}
\vspace{2cm}\\
Karol Ko\l odziej$^{\rm a,}$\footnote{
        On leave from the Institute of Physics, University of Silesia,
        PL--40007 Katowice, Poland},
Arnd~Leike$^{\rm a}$
        and
Reinhold~R\"uckl$^{\rm a,b}$\vspace{0.5cm}\\
$\nobody^{\rm a}${\small\it
Sektion Physik der Universit\"at M\"unchen,\\
 Theresienstr. 37, D--80333 M\"unchen, FRG}\\
$\nobody^{\rm b}${\small\it
Max-Planck-Institut f\"ur Physik, Werner-Heisenberg-Institut,\\
F\"ohringer Ring~6, D-80805 M\"unchen, FRG}
\vspace*{4cm}\\
{\bf Abstract}
\end{center}
{\small
Using photon-photon collisions as a particularly transparent study
case we investigate the production mechanisms for $B_c$ mesons. In
nonrelativistic approximation and to $O(\alpha^2\alpha_s^2)$ it is
shown that recombination of $\bar{b}$- and $c$-quarks dominates by
far over $\bar{b}$ and $c$ fragmentation. This dominance persists
up to the highest accessible transverse momenta and leads to
distributions in energy which differ completely from the spectra
expected on the basis of the fragmentation functions.
For processes in which a $b \bar{b}$-pair is radiated from a primary
$c$-quark, the fragmentation description is found to be inadequate.
We anticipate important implications of these results for hadronic
production of heavy quark resonances. Using realistic photon spectra
we predict two-photon production rates for $B_c$ and $B_c^*$ at
present and future $e^+e^-$-machines.
}
\vfill\newpage
%
\section{Introduction}

Since the top quark seems to be too short-lived for quarkonium-like
resonances to form, the $\bar{b}c$-bound states are most likely the
only narrow heavy quark resonances besides the well-known
charmonium and bottomonium systems. Till now these so-called $B_c$
mesons have not yet been observed. However, they are expected to
be in experimental reach with some luck already at LEP, more likely
at the TEVATRON, but most certainly at the LHC. In addition,
also the HERA-B experiment may get a glimpse on these very
interesting bound states.

Because of flavour conservation in strong and electromagnetic
processes, the $B_c$ ground state can only decay weakly unlike
the $\eta_c$ and $\eta_b$. This fact provides unique possibilities
to investigate important aspects of strong and weak interactions
and their interplay. Indeed, the nonrelativistic nature of these
systems allows to calculate also genuinely nonperturbative
quantities such as total  production cross sections,
fragmentation functions, decay constants, and transition
form factors. For this reason, the $B_c$ system can serve as
a testing ground for quantitative approaches to confinement
problems and our understanding of soft hadron physics.

So far, the estimates only include the contributions of lowest
order in the relative momentum of the constituent quarks and
 in the strong coupling constant. Moreover, they are subject to
numerical uncertainties in the bound state wave
functions, relevant quark masses and coupling constants.
Independently of that, the calculations have not yet reached
complete agreement on the hadronic production cross sections.
Also, there are different opinions on the relative importance of
recombination and fragmentation mechanisms, such as
$\bar{b}c \ra B_c$ and $\bar{b} \ra B_c \bar{c}$, respectively.

In this paper, we take up the latter question and clarify
the issue of recombination versus fragmentation. In order to
liberate our study from unnecessary complications, we consider
photon-photon instead of gluon-gluon scattering
which is the dominant subprocess in hadronic collisions.
The two reactions have many features in common including the
competition of recombination and fragmentation processes.
On the other hand, the absence of coloured
quanta in the initial state of $\gamma\gamma$-processes drastically
reduces the number of Feynman diagrams at a given order and
makes the calculations much more transparent. We realize, however,
that the nontrivial colour coefficients in $gg$-processes
somewhat change the relative weight of the different mechanisms
in comparison to $\gamma \gamma$-scattering.

Preliminary results of our study were already reported in ref.~\cite{lr}.
However, there the focus was more on the prospects of observing $B_c$ mesons
in $e^+e^-$-annihilation. Here, we complete our analysis of $B_c$
production in $\gamma\gamma$ collisions and also correct a normalization
error in the fragmentation estimate given in ref.~\cite{lr}.
Furthermore, we show the relevant differential distributions and
point out the features which distinguish recombination from
fragmentation.
In addition, we examine in detail the theoretical validity of the
factorized description in terms of heavy quark production,
$\gamma\gamma \ra c \bar{c}$ and $b \bar{b}$, and subsequent
fragmentation, $c \ra B_c b$ and $\bar{b} \ra B_c \bar{c}$, respectively.
Finally, we present predictions on the total
cross section for $B_c$ and $B_c^*$ production in collisions of high
energy photon beams which may be obtained by back-scattering of low
energy, high intensity laser beams at future Linear Colliders,
and in collisions of softer bremsstrahlung photons.

When our work was near to completion, a related study appeared in
ref.~\cite{bls}. If we use the same parameters
we reproduce the total cross sections within the Monte Carlo
uncertainties, except for the highest $\gamma \gamma$ c.m. energy
of 100 GeV considered in ref.~\cite{bls}, where we disagree by ten per cent.
Also the differential
distributions presented in ref.~\cite{bls} show in general the same
behaviour as ours.
%
\section{Classification of processes and calculation}
In $\gamma\gamma$ collisions, $B_c$ mesons are dominantly produced in
association with a $b$- and $c$-quark jet,
$$\gamma(k_1)+\gamma(k_2) \ra B_c(P) + b(p_b)+ \bar{c}(p_c).$$
Here, the particle momenta are indicated in parentheses.
To lowest order in the electromagnetic and strong coupling constants
$\alpha$ and $\alpha_s$, respectively, the above process is described
by twenty Feynman diagrams.
These can be classified in different topologies as indicated in Fig.~1.
The diagrams belonging to set (I) can further be distinguished by the
flavour of the primary quark line coupled to the photons. This line carries
either the bottom (I$_b$) or the charm (I$_c$) flavour.
The complete subsets of six
diagrams with primary $b$- and $c$-quarks are found by
interchanging the initial photons as required by Bose symmetry.
These subsets are individually gauge invariant.
The remaining eight diagrams belonging to set (II) are characterized by
direct coupling of the initial photons to the two different
quark flavours. The complete set (II) is obtained from the representative
diagrams shown in Fig.~1
by interchanging independently the photons and quark flavours. Of course,
also set (II) is by itself gauge invariant.

\vspace*{-0.5cm}
\begin{center}
\begin{minipage}[t]{7.8cm} {
\begin{center}
\hspace{-1.7cm}
\mbox{
\epsfysize=7.0cm
\epsffile[0 0 500 500]{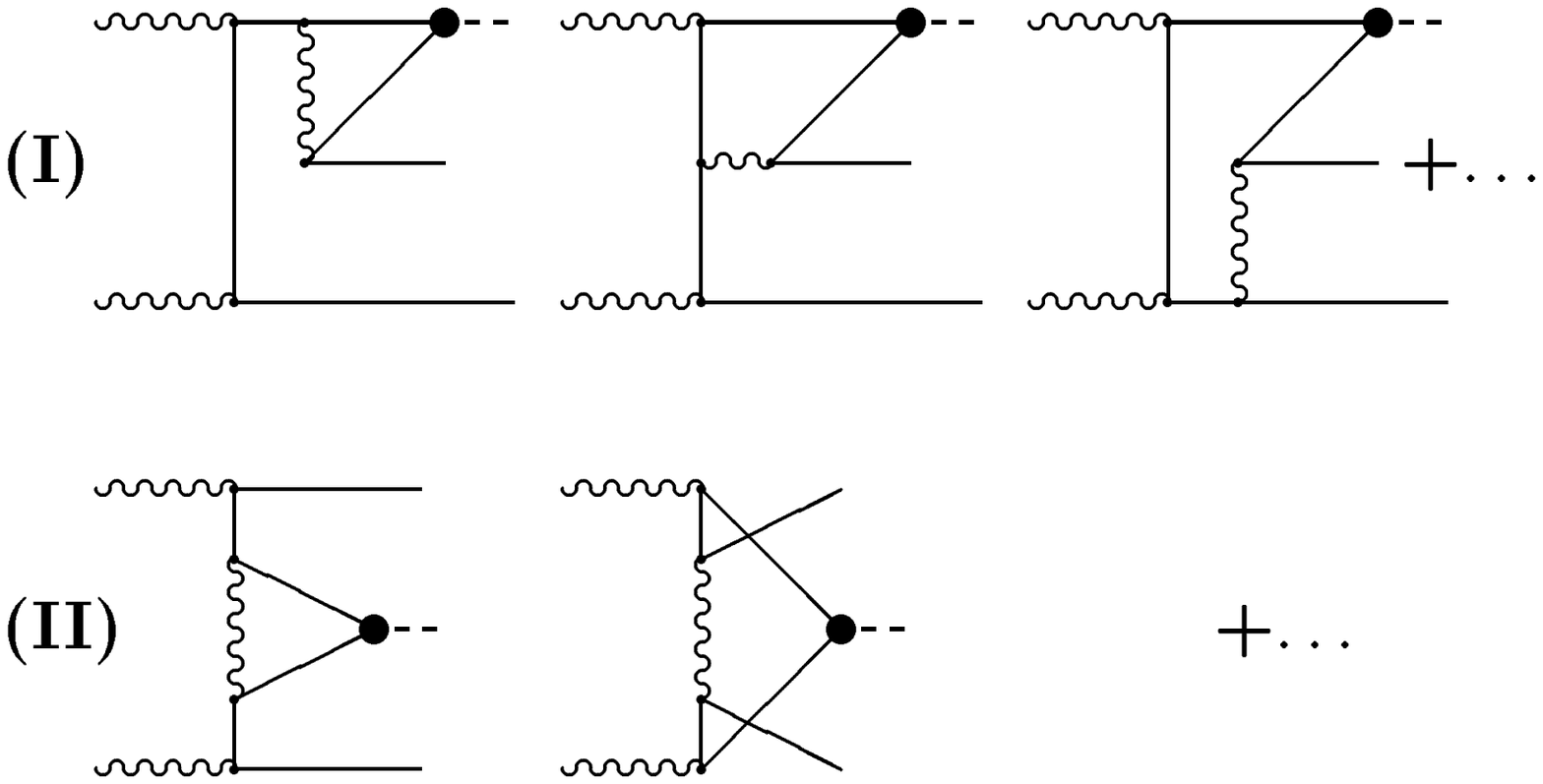}
}
\end{center}
\vspace*{-2.0cm}
\noindent
{\small\bf Fig.~1. }{\small\it
Different topologies of the lowest-order Feynman diagrams contributing
to $\gamma \gamma \rightarrow B_c b \bar{c}$.
}
}\end{minipage}
\end{center}
\vspace{0.5cm}

More physically, the three subsets of diagrams discussed above can be
interpreted as describing different production mechanisms. The subset
(I$_b$) with primary $b$-quarks may be associated with $b \bar b$
production and $\bar b$-quark fragmentation $\bar b\rightarrow B_c\bar c$
in accordance with the first diagram
of Fig.~1 (I).
Similarly, one may be tempted to attribute the subset
(I$_c$) to $c \bar c$ production followed by
$c$-quark fragmentation $c\rightarrow B_c b$.
However, while the fragmentation description is applicable to the
process (I$_b$), it does not hold for (I$_c$) as will be shown
later. In contrast to the class (I) diagrams, the class (II) diagrams
are characterized by direct production of a $b \bar b$ and
$c \bar c$ pair from the initial photons and recombination of the
$\bar b$- and $c$-quark into
a $B_c$ meson, $\bar b c\rightarrow B_c$. Obviously, the production
amplitudes for the three
different mechanisms are proportional to different combinations of the
electric quark charges, namely $Q_b^2=1/9,\ Q_c^2=4/9$ and $Q_bQ_c=-2/9$,
respectively. As one of the main results, we will show later that
the recombination mechanism (II) dominates, while $b \bar b$ radiation from a
primary $c$-quark (I$_c$) gives a small contribution and
fragmentation of a primary $\bar b$-quark (I$_b$) is completely
negligible.

We begin with a brief description of technical details of the calculation of
amplitudes and cross sections.
For definiteness and because of their particular importance, we concentrate on
the pseudoscalar and vector ground states, $B_c$ and $B_c^*$, respectively.
Because of the
nonrelativistic nature of these systems, the relative momentum $p$
of the $\bar b$ and $c$ constituents and their binding energy are
expected to be small in comparison to the quark masses.
It is then reasonable to expand the amplitudes in $p$, and to
keep only the lowest nonvanishing term \cite{GKPR}.
In this approximation, the $\bar b$- and $c$-quark are on mass shell and move
together with equal velocity. The following relations hold:
\begin{equation}
\label{bound}
M(B_c)=M(B_c^*)=M=m_c+m_b,\ \ \ p_c = \frac{m_c}{M} P,
\ \ \ p_{\bar{b}} = \frac{m_b}{M} P.
\end{equation}
Moreover, the amplitudes for the production of $S$--waves reduce to hard
scattering
amplitudes multiplied by the $S$--wave function $\Psi(0)$ at the origin.
For $P$--waves one obtains a similar product involving the derivative of the
$P$--wave function. Using the definitions
\bq
<0|\bar b\gamma_\mu\gamma_5 c|B_c(P)>\  =\  if_{B_c}P_\mu
\eq
and
\bq
<0|\bar b\gamma_\mu c|B_c^*(P)>\  =\  Mf_{B_c^*}\epsilon_\mu
\eq
one can substitute the nonrelativistic wave function parameter $\Psi(0)$
by the corresponding decay constants
\bq
f_{B_c} = f_{B_c^{*}} = \sqrt{\frac{12}{M}}|\Psi(0)|.
\eq
As a simple rule, in the Feynman diagrams of Fig.~1 the transition from
open quark production, $\gamma\gamma\rightarrow \bar b c b \bar c$, to
$B_c^{(*)}$ bound state production is achieved by making the following
replacement \cite{GKPR}:
\begin{equation}
\label{nonrela}
v(p_{\bar b})\bar u(p_c) = \frac{f_{B_c^{(*)}}}{\sqrt{48}}(P\!\!\!\!/ - M)
                           \Pi_{SS_Z}\; .
\end{equation}
Here, $v(p_{\bar b})$ and $\bar u(p_c)$ denote the Dirac spinors of the quarks
forming the $B_c^{(*)}$ bound states, and
 $\Pi_{SS_Z}$ is the appropriate spin projector, $\Pi_{00}=\gamma_5$ for
pseudoscalars and $\Pi_{1S_Z}=\rlap/\epsilon$ for vectors.
In Fig.~1, this formal substitution is indicated by the dark dot. Obviously,
because of the neglect of the relative momentum of the constituent quarks,
the Feynman amplitudes involve no loop in momentum space, but only traces
of Dirac and colour  matrices. Note that the colour structure is not
accounted for in eq.~(\ref{nonrela}). All diagrams of Fig.~1 have the
same colour factor ${4 \over 3}$ which has been included in the overall
normalization of the cross sections.

The squared matrix element is obtained by two independent methods.
One calculation is based on the traditional trace technique. The
necessary traces are calculated using the symbolic manipulation
program {\tt FORM}~\cite{form}. The resulting expressions are automatically
implemented into a {\tt FORTRAN} program.
The second calculation is based on a generalization of the method
described in detail in ref.~\cite{KZ} for bosonic final states in
$e^+e^-$ scattering. Here, the polarized matrix elements are calculated
in a fixed Lorentz frame using the Weyl representation of the Dirac matrices
and spinors together with real polarization vectors.
The matrix element is reduced analytically to a linear combination of
products of the $2\times 2$ Pauli matrices. These products are computed
numerically. The use of real photon polarization vectors allows to reduce
the number of elements of the matrices
which have to be calculated by a factor of two.
The polarized matrix elements are obtained
by sandwiching these matrices between $2\times 2$ Pauli spinors.
As the last step, one
sums over polarizations.
The matrix elements squared obtained by the two methods agree to ten digits.

The phase space integration is also performed independently using two
different routines. Before employing the Monte Carlo integration
routine {\tt VEGAS}~\cite{vegas} the strongest peaks of the matrix
elements were eliminated
by introducing new integration variables.
After the integration all differential distributions and cross
sections agree within the Monte Carlo errors.
\section{Results}
%
For numerical purposes and all subsequent figures, we choose the
following values of the parameters:
\begin{eqnarray}
\label{input}
m_b  = 4.8\ {\rm GeV}, \ \ \ m_c  = 1.5\ {\rm GeV},\ \ \ \alpha_s  = 0.2,
\ \ \ \alpha = 1/129, \ \ \ f_{B_c} = f_{B_c^*} = 0.4\ {\rm GeV}
\ {\rm\cite{RR}}.
\end{eqnarray}
In Fig.~2 we compare the cross section $\sigma({\rm I}_b+{\rm I}_c+{\rm II})$
for $\gamma\gamma\rightarrow B_c b \bar c$ derived from the complete set
of $O(\alpha ^2 \alpha ^2_s)$ diagrams indicated in Fig.~1 with the cross
sections $\sigma({\rm I}_b)$ and $\sigma({\rm I}_c)$ resulting
 from the gauge invariant subset (I) with primary $b$- and $c$-quarks,
respectively.
Also shown are the results obtained by simply multiplying the cross sections
$\sigma_{b\bar b}$ and $\sigma_{c\bar c}$ for open flavour production with the
respective fragmentation probabilities $P(\bar b\rightarrow B_c\bar c)$ and
$P(c\rightarrow B_c b)$ \cite{dz} . These probabilities have been calculated
separately in the same nonrelativistic approximation as described in
section~2.

As can be seen, the total production cross section
is dominated by the recombination mechanism except very near to the
production threshold, where also fragmentation, or more precisely radiation,
processes give significant
contributions. Moreover, it is more efficient to produce a pair of $c$-quarks
and to radiate from them a $b\bar b$-pair in order
to form a $B_c$ meson than to produce a pair of $b$-quarks and to radiate a
$c\bar c$-pair. The reason is the electric charge of the primary quarks
which yields a factor 16 in favour of $c$-quarks.
Finally, the factorized description in terms of open flavour production,
$\gamma\gamma\rightarrow b\bar b$ and $\gamma\gamma\rightarrow c\bar c$,
followed by the fragmentation processes, $\bar b\rightarrow B_c\bar c$ and
$ c\rightarrow B_c b$, respectively, provides a good approximation to
the class (I) processes only when the primary quark is a $b$-quark,
but not when
it is a $c$-quark. These findings are in sharp contrast to $B_c$ production
in $e^+e^-$-annihilation \cite{EE, fragee}. There, the recombination
mechanism is
absent and the radiation of a $b \bar b$ pair from primary $c$-quarks
is negligible. Furthermore, the
fragmentation picture works perfectly no matter whether the fragmenting
quark is the heavier or the lighter one.

\vspace*{-0.5cm}
\begin{center}
\begin{minipage}[t]{7.8cm} {
\begin{center}
\hspace{-1.7cm}
\mbox{
\epsfysize=7.0cm
\epsffile[0 0 500 500]{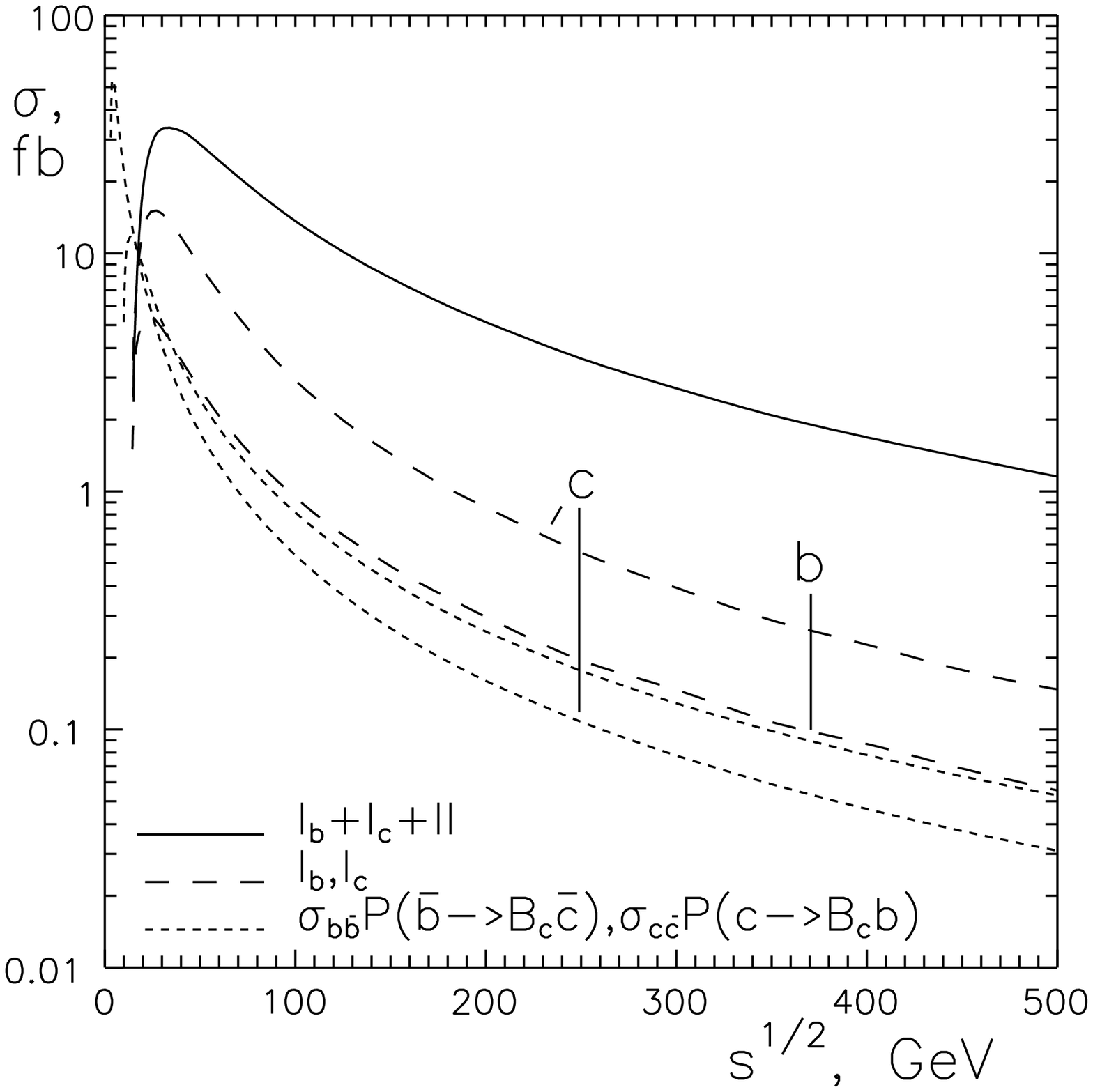}
}
\end{center}
\noindent
{\small\bf Fig.~2. }{\small\it
Integrated cross sections for $\gamma\gamma \rightarrow B_c b \bar{c}$
versus the c.m. energy. The different calculations are explained in the text.
The production mechanisms are as classified in Fig.~1.
}
}\end{minipage}
\end{center}
\vspace*{0.5cm}

We have investigated the break-down of the fragmentation description for the
class (I) processes in $\gamma\gamma$-scattering in some more detail.
In energy range considered, this approximation is indeed only applicable
when the fragmenting quark is
appreciably heavier than the quark flavour produced via gluon radiation.
Such a favourable case is illustrated in Fig.~2 by $\sigma({\rm I}_b)$ and
$\sigma_{b\bar b}P(\bar b\rightarrow B_c\bar c)$ which agree nicely.
However, already for equally heavy quarks, a case realized in the analogous
process $\gamma\gamma\rightarrow J/\psi c\bar c$, we find considerable
disagreement. To be more definite,
$\sigma(\gamma\gamma \ra J/\psi\bar c c)/\sigma_{c\bar c} 2
P(\bar c \rightarrow J/\psi\bar c ) \approx 2.1\ (1.4)$\
at $\sqrt s =20\ (500)$\ GeV.
When the fragmenting quark is the lighter one, the fragmentation
description fails completely as demonstrated by $\sigma({\rm I}_c)$ and
$\sigma_{c\bar c}P(c\rightarrow B_c b)$ in Fig.~2. It is the second diagram of
Fig.~1 (I) which seems to be responsible for the break-down of factorization.
This diagram is absent in $e^+e^-$ annihilation, but it is necessary in
$\gamma\gamma$-scattering for gauge invariance.

Further features of the different production mechanisms and theoretical
descriptions
can be learned from differential distributions of the $B_c$ mesons.
Fig.~3 shows the distributions in the scaled energy variable
$z=2E_{B_c}/\sqrt{s}$ obtained from the complete set of
$O(\alpha^2 \alpha_s^2)$ diagrams and from the gauge-invariant subsets
(I$_b$) and (I$_c$) alone.
Most notably, the dominant recombination processes
populate the soft end of the spectrum, while the class (I) processes give rise
to a characteristic high energy tail. The hard component (I$_b$) is as
expected from $b$-quark fragmentation. In fact, the energy distribution
predicted from the diagrams of class (I$_b$)
is practically indistinguishable from the corresponding
fragmentation function $D_{\bar b\rightarrow B_c}(z)$ \cite{dz, fragee}.
However, for the component (I$_c$) the analogous distributions do not
coincide. The actual energy spectrum derived from the diagrams (I$_c$)
is considerably harder than the fragmentation function $D_{c \ra B_c}(z)$
\cite{dz, fragee}. The above assertions are illustrated more clearly
in Fig.~4 where we plot the individual energy spectra generated by the
processes (I$_b$), (I$_c$) and (II) in comparison to the fragmentation
functions $D_{\bar b \ra B_c}(z)$ and $D_{c \ra B_c}(z)$. Here, all
distributions are normalized to unity.
\ \vspace{1cm}\\
\begin{minipage}[t]{7.8cm} {
\begin{center}
\hspace{-1.7cm}
\mbox{
\epsfysize=7.0cm
\epsffile[0 0 500 500]{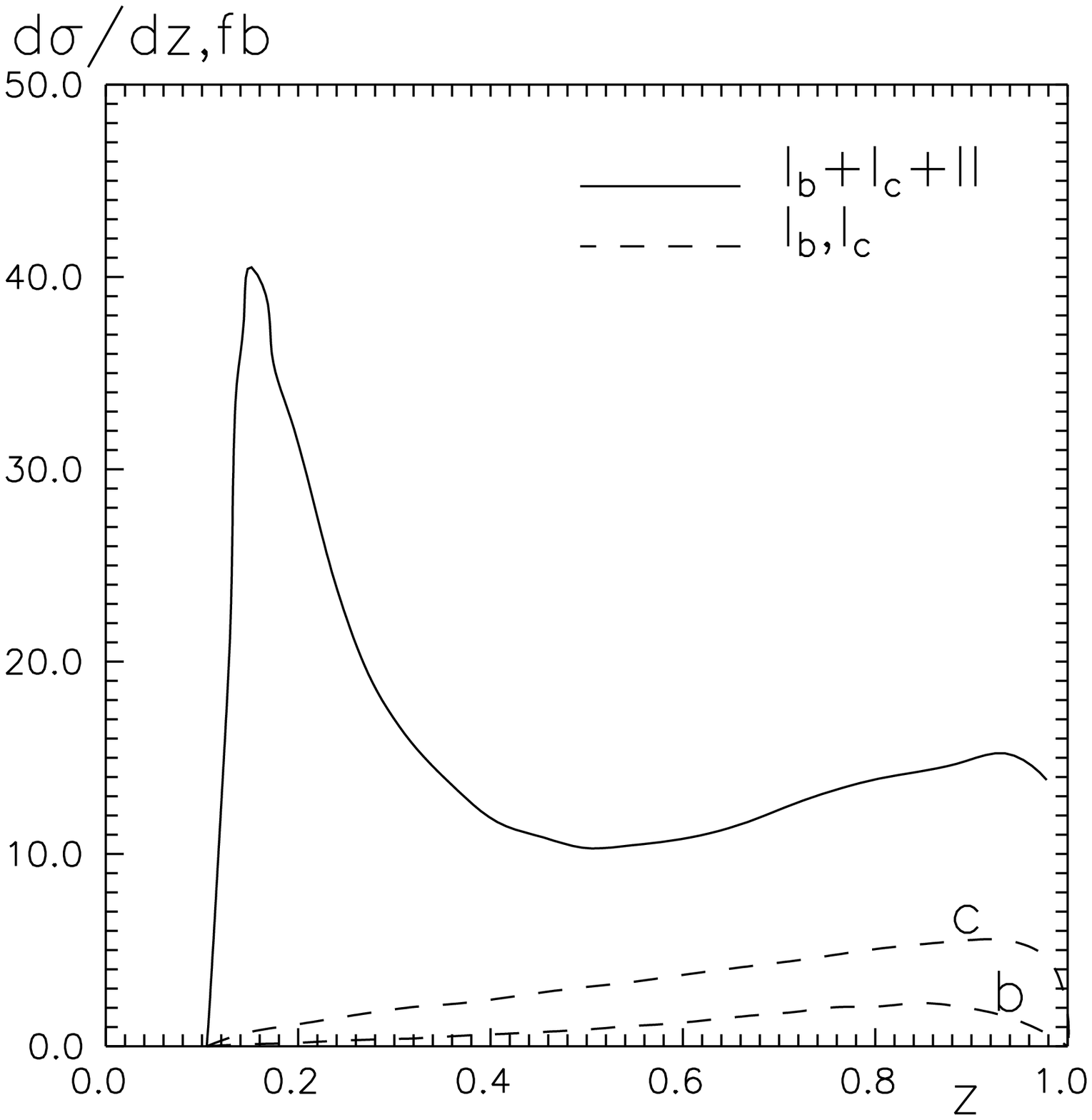}
}
\end{center}
\vspace*{-0.5cm}
\noindent
{\small\bf Fig.~3. }{\small\it
Energy distributions of $B_c$ mesons at $\sqrt{s}=100$\ GeV
with all processes of Fig.~1 included and from the radiative mechanisms
(I$_b$) and (I$_c$) alone.
}
}\end{minipage}
\hspace{0.5cm}
\begin{minipage}[t]{7.8cm} {
\begin{center}
\hspace{-1.7cm}
\mbox{
\epsfysize=7.0cm
\epsffile[0 0 500 500]{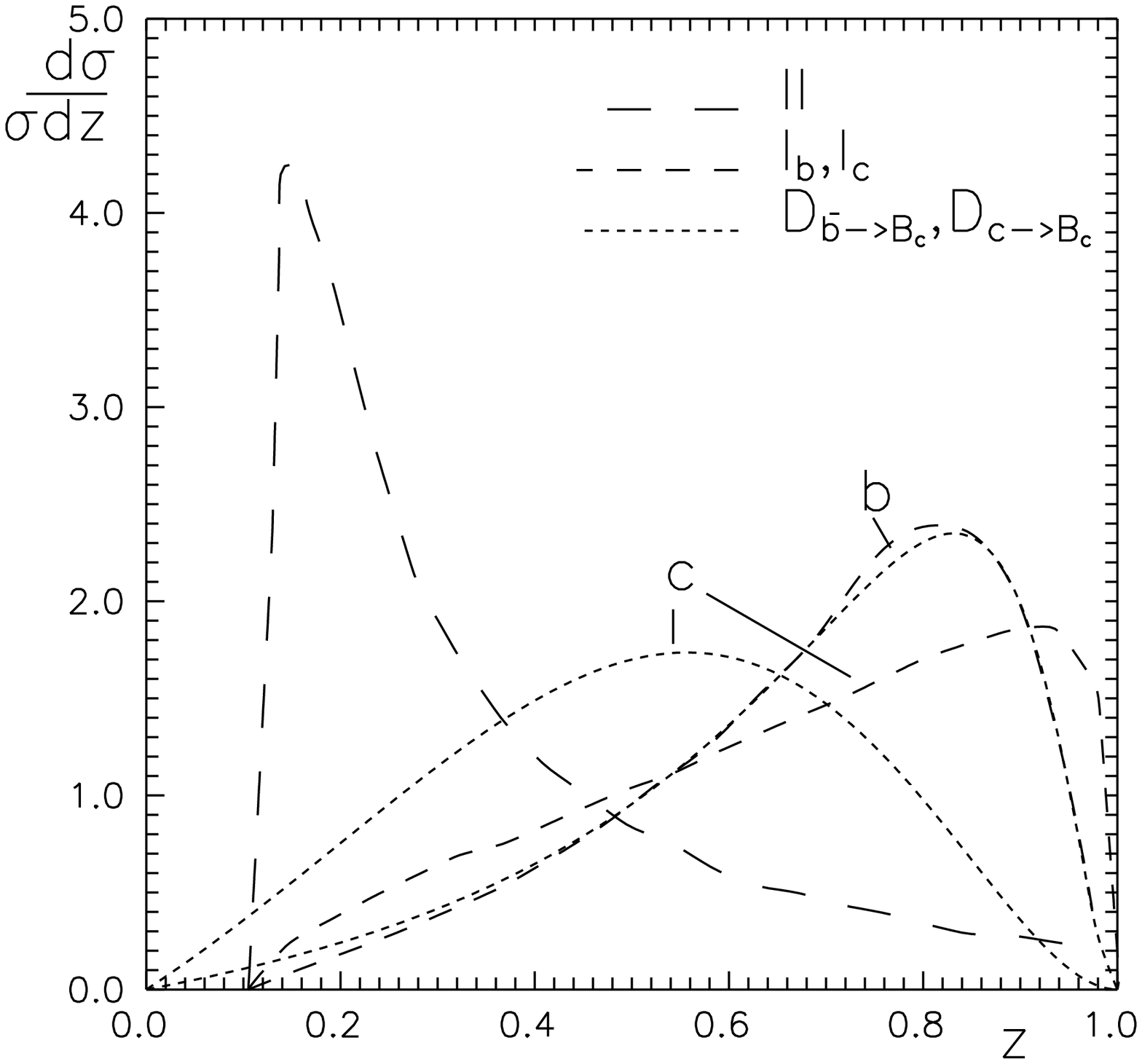}
}
\end{center}
\vspace*{-0.5cm}
\noindent
{\small\bf Fig.~4. }{\small\it
Normalized energy spectra of $B_c$ mesons at $\sqrt{s}=100$\ GeV
generated by the mechanisms (II), (I$_b$) and (I$_c$) of Fig.~1
in comparison to the heavy quark fragmentation functions.
}
}\end{minipage}
\vspace*{0.5cm}

Interesting observations can also be made from the
$p_{T}$ distributions depicted in Fig.~5.
The factorized description in terms of $b \bar b$ production and
$\bar b$-fragmentation and the corresponding diagrams (I$_b$) of Fig.~1
yield distributions which are very similar in shape and normalization
except in the low-$p_T$ region. As far as the shape is concerned this
is even true for the analogous processes with primary $c$-quarks.
However, the $c$-quark fragmentation picture fails to reproduce
the correct magnitude of the cross section for the process (I$_c$),
a problem already noticed in Fig.~2.
We also see that the radiation of a $c \bar c$-pair from a $b$-quark
leads to a harder $p_T$ distribution for the $B_c$ bound states than
the radiation of a $b \bar b$-pair from $c$-quarks.
Finally, the most important observation
is that the recombination mechanism dominates $B_c$ production not only at
low $p_T$, as one could have expected, but also in the high-$p_T$ region up
to the kinematic limit. In other words, the familiar description of high-$p_T$
hadron production in terms of the production and fragmentation of quarks with
large transverse momenta is inadequate for single $B_c$ production
in $\gamma\gamma$-scattering and similar processes.

For completeness, we mention without showing figures that the angular
distributions resulting from the different mechanisms considered above are all
quite similar in shape. The distributions are sharply peaked in forward and
backward direction, with the distribution predicted by the complete set of
diagrams of Fig.~1 being slightly flatter than the distributions derived from
subsets (I$_b$) and (I$_c$), and from $b$- and $c$-quark fragmentation.

So far we have concentrated on the problem of identifying the dominant
production mechanisms and examining different theoretical descriptions. For
this purpose, it is sufficient to consider photon-photon collisions at
fixed energies. Now we want to make predictions on the production rates of
$B_c$ mesons in photon-photon collisions at $e^+e^-$-machines in the LEP
energy range and beyond.
To this end we fold the total
cross sections for $\gamma \gamma \ra B_c^{(*)} b \bar{c}$
with realistic photon spectra that is the photon
spectrum obtained by Compton back-scattering of high intensity laser light
on $e^{\pm}$ beams \cite{Ginzburg}, or the Weizs\"acker--Williams
bremsstrahlung spectrum \cite{WW}. Our results are shown in Fig.~6, where
we have plotted the convoluted cross sections for production of the
pseudoscalar and vector ground states versus the $e^+e^-$ centre-of-mass
energy. For reference we also show the unfolded cross
sections for the subprocesses $\gamma \gamma \ra B_c^{(*)} b \bar{c}$
as functions of the $\gamma \gamma$ centre-of-mass energy.
As expected from the shape of the Compton spectrum which has a long
tail towards soft photons, and the shape of the $\gamma\gamma$-cross section
which peaks just above threshold,
the convolution
increases the cross sections substantially for energies above 100\,GeV.
At a 500 GeV linear collider and for an integrated luminosity of 10 fb$^{-1}$,
one can produce about 100 $B_c$ and 400 $B_c^*$ mesons.
In contrast, the yield of $B_c$ mesons from bremsstrahlung photons is very
small at LEP energies, but increases logarithmically with energy. While
at LEP 2 the production rate is still too low to lead to observable signals,
in the TeV energy range bremsstrahlung photons become competitive with
back-scattered laser photons in producing $B_c$ mesons.

\ \vspace{1cm}\\
\begin{minipage}[t]{7.8cm} {
\begin{center}
\hspace{-1.7cm}
\mbox{
\epsfysize=7.0cm
\epsffile[0 0 500 500]{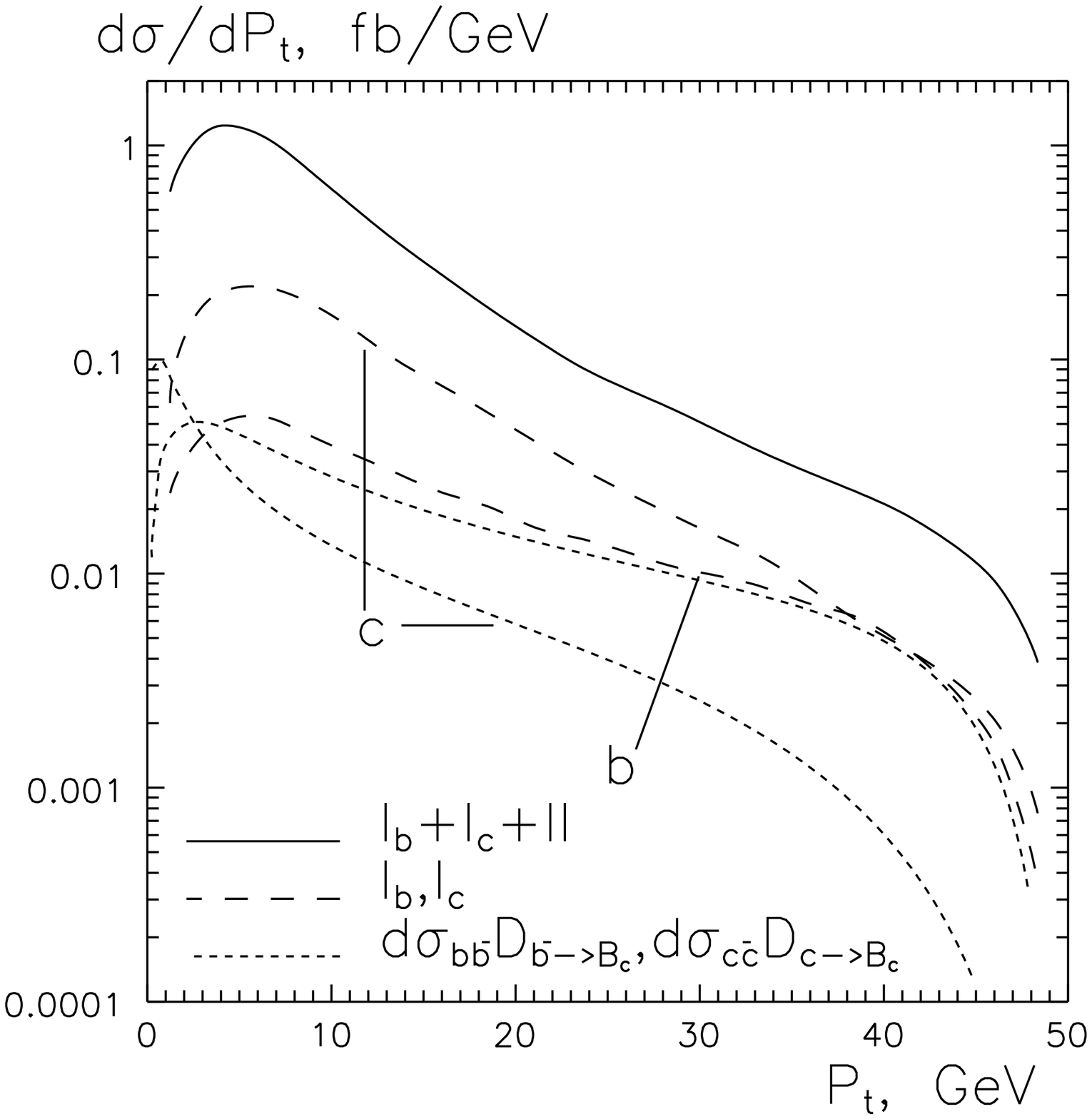}%
}
\end{center}
\noindent
{\small\bf Fig.~5. }{\small\it
The $B_c$ transverse momentum distributions at $\sqrt{s}=100$\ GeV
for the different mechanisms and descriptions considered in Fig.~2.
}
}\end{minipage}
\hspace{0.5cm}
\begin{minipage}[t]{7.8cm} {
\begin{center}
\hspace{-1.7cm}
\mbox{
\epsfysize=7.0cm
\epsffile[0 0 500 500]{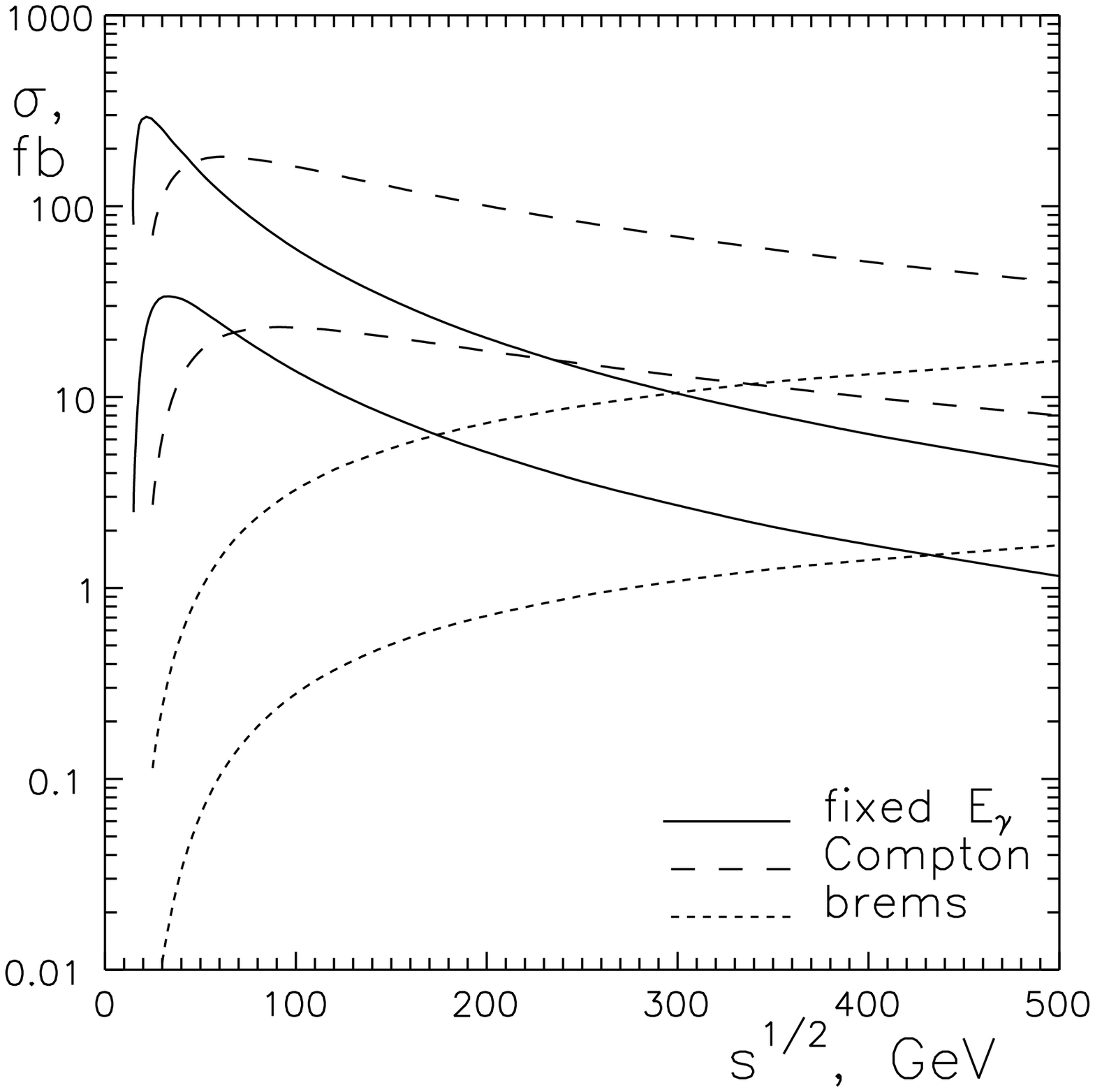}%
}
\end{center}
\noindent
{\small\bf Fig.~6. }{\small\it
Total production cross sections for $B_c^*$ (upper curves) and
$B_c$ (lower curves) versus the $\gamma \gamma$ c.m. energy (full),
and convoluted with the Compton spectra of back-scattered laser light
(long-dashed) and with bremsstrahlung spectra (short-dashed)
versus the $e^+e^-$ c.m. energy.
}
}\end{minipage}
\vspace*{0.5cm}
\section{Summary}
We have studied the production of $B_c$ mesons in photon-photon
scattering. Our main results can be summarized as follows. The
dominant production mechanism proceeds via direct coupling of
a $b \bar b$ and $c \bar c$ pair to the initial photons and recombination
of the $\bar b$- and $c$-quark into the $B_c$ bound state.
The radiative process
where a $c \bar c$-pair is produced via gluon emission from a primary
$b$-quark plays a negligible role. It is well described by $b$-quark
production and fragmentation, $\gamma\gamma \ra b \bar b$,
$\bar b \ra B_c \bar c$. The analogous process where a $b \bar b$-pair
is emitted from a primary $c$-quark is
enhanced by the charge ratio $Q_c^4/Q_b^4$, but it still plays a marginal
role. Here, the fragmentation description $\gamma\gamma \ra
c \bar c$, $c \ra B_c b$ does not hold.

Given the dominance of recombination even at large transverse momenta
up to the kinematic limit we conclude that the usual hard scattering
formalism in terms of high-$p_T$ quark production and subsequent
fragmentation cannot be applied to $\gamma \gamma \ra B_c X$.
This also puts serious doubts on the calculations of hadronic $B_c$
\cite{PP2} and $J/\psi$ \cite{Roy} production based on $b \bar b$
and $c \bar c$ production via gluon-gluon fusion followed by heavy
quark fragmentation. Although the replacement of the electric
charges by colour charges when going from $\gamma \gamma $- to
$g g$-processes changes the relative weights of the production
mechanisms considered in this paper, we suspect the fragmentation
description to underestimate the yields of heavy bound states
in the experimentally accessible, high-$p_T$ regime.

The $B_c^{(*)}$ production rates predicted for $\gamma \gamma$ collisions
and realistic photon energy distributions show that these processes
are below the observability limit in $e^+e^-$-collisions at LEP
energies. However, at linear colliders in the TeV energy range,
two-photon production of $B_c$ mesons may come into experimental
reach. These general prospects are little affected by the sizeable
uncertainties in the values of parameters quoted in eq.~(\ref{input}),
and the unknown higher order perturbative, nonperturbative and
relativistic corrections.
%

\end{document}